\documentclass[12pt]{iopart}
\usepackage{bm,graphicx,amssymb,subfig,psfrag}

\usepackage{graphicx}

\usepackage{cite}
\usepackage{color} 
\usepackage[plain]{algorithm}
\usepackage{algorithmic}

\usepackage{hyperref}

\newcommand{\puntini}{\ensuremath{..}}
\newcommand{\cM}{\ensuremath{{\mathcal M}}}
\newcommand{\cMc}{\ensuremath{{\mathcal M}^c}}
\newcommand{\cL}{\ensuremath{{\mathcal L}}}
\newcommand{\cLc}{\ensuremath{{\mathcal L}^c}}
\newcommand{\bM}{\ensuremath{{\mathrm b}_\cM}}
\newcommand{\eM}{\ensuremath{{\mathrm e}_\cM}}
\newcommand{\iset}{\ensuremath{i\in [1\puntini 4]}}
\newcommand{\sw}{\ensuremath{\cL_{\mathrm sw}}}
\newcommand{\Xcl}{\textsf{Xclean}}

\begin{document}
\title[Computing the HOMFLY Polynomial of Proteins]{A Topological Framework for the Computation of the HOMFLY Polynomial and its Application to Proteins}

\author{Federico Comoglio$^{1,2}$ and Maurizio Rinaldi$^{1}$}
\address{\bf{1} Department of Chemical, Food, Pharmaceutical and Pharmacological Sciences (DiSCAFF), University of Piemonte Orientale ``Amedeo Avogadro'', Novara, Italy. \bf{2} Current address: Department of Biosystems Science and Engineering (D-BSSE), ETH Zurich, Basel, Switzerland}
\ead{rinaldi@pharm.unipmn.it}
\begin{abstract}
Polymers can be modeled as open polygonal paths and their closure generates knots. Knotted proteins detection is currently achieved via high-throughput methods based on a common framework insensitive to the handedness of knots.\\
Here we propose a topological framework for the computation of the HOMFLY polynomial, an handedness-sensitive invariant. %, via the geometric construction of Conway skein triples. 
Our approach couples a multi-component reduction scheme with the polynomial computation. After validation on tabulated knots and links the framework was applied to the entire Protein Data Bank along with a set of selected topological checks that allowed to discard artificially entangled structures. This led to an up-to-date table of knotted proteins that also includes two newly detected right-handed trefoil knots in recently deposited protein structures.\\
The application range of our framework is not limited to proteins and it can be extended to the topological analysis of biological and synthetic polymers and more generally to arbitrary polygonal paths.
\end{abstract}
{\scriptsize \textbf{Citation: } Comoglio F, Rinaldi M (2011) A Topological Framework for the Computation of the HOMFLY Polynomial and Its Application to Proteins. PLoS ONE 6(4): e18693. doi:\href{http://www.plosone.org/article/info%3Adoi%2F10.1371%2Fjournal.pone.0018693?utm_source=feedburner&utm_medium=feed&utm_campaign=Feed%3A+plosone%2FMathematics+(PLoS+ONE+Alerts%3A+Mathematics)}{10.1371/journal.pone.0018693}}

\section{Introduction}
\begin{figure}
\includegraphics[width=0.6\textwidth]{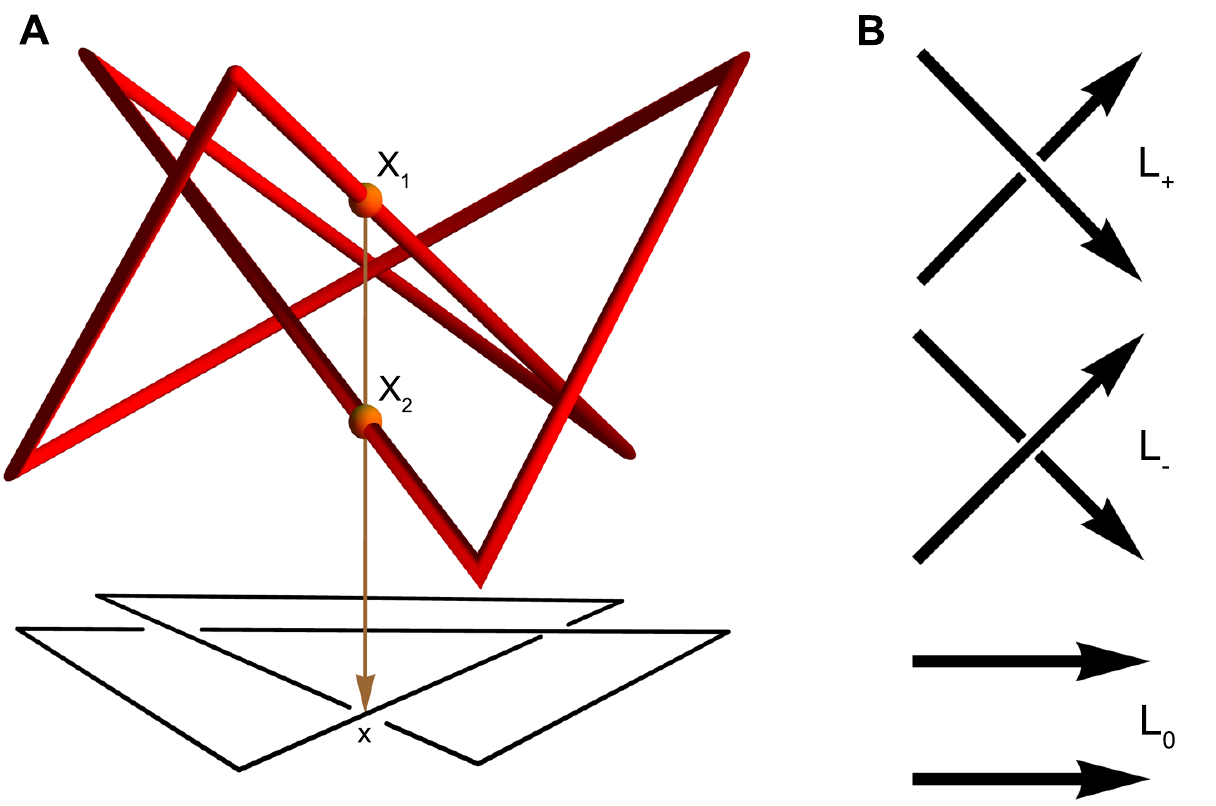}
\caption{
{\bf A knot diagram and illustration of the Conway skein triple} (A) Three dimensional polygonal representation of the trefoil knot (in red) and its planar diagram (in black). Two red spheres on the knot mark the 3D points $X_1$ and $X_2$ projecting down to $\mathrm{x}$ on the planar diagram along the brown arrow. (B) The Conway skein triple is composed of three oriented diagrams that are the same outside a small region, where they look like the illustrated $L_+$, $L_-$ and $L_0$. To define the oriented sign of a crossing, approach it along the underpass in the direction of the orientation: if the overpass orientation runs from left to right, the oriented sign is $+1$, $-1$ otherwise.}
\label{Figure_1}
\end{figure}

The topological study of biological polymers has led to important insights into their structural properties and evolution~\cite{Sumners,TaylorIII}. 
From a topological point of view polymers can be naturally modeled as sequences of 3D points, i.e.~open polygonal paths. Their closure generates classical objects in topology called knots. The simplest knot is the trefoil knot, illustrated in Figure~\ref{Figure_1}A. The characterization of knotted proteins, due to their close structure-function relationship and reproducible entangled folding, is a subject of increasing interest in both experimental and computational biology.\\

\begin{figure}
\includegraphics[width=0.6\textwidth]{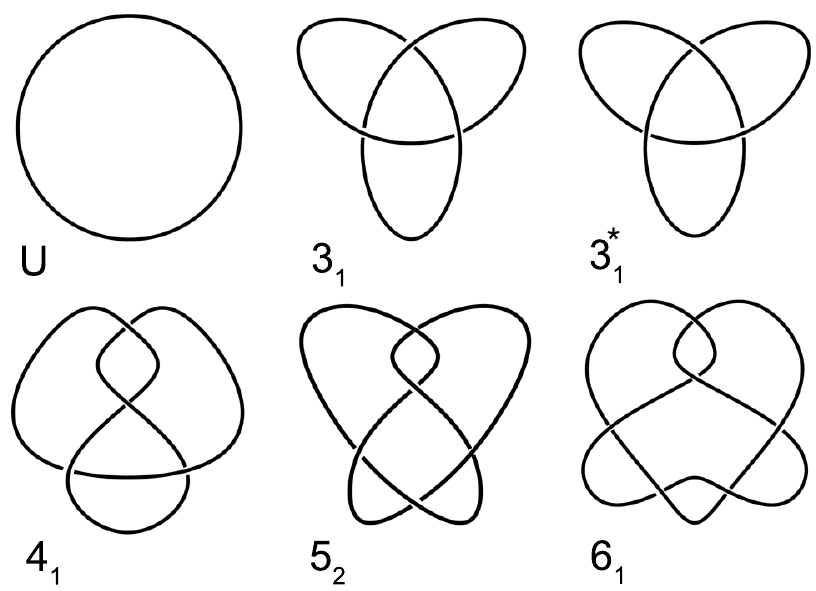}
\caption{
{\bf Knots met in proteins}  Illustration of the knots found in proteins, labeled according to Rolfsen names. U: the simplest knot, the unknot. $3_1$: the trefoil knot and its mirror image, denoted by the $*$, has three crossings. $4_1$: the figure-eight knot is the only knot with four crossings. $5_2$: the three-twist knot has five crossings. $6_1$: the Stevedore's knot, the most complex knot detected in proteins.}
\label{Figure_2}
\end{figure}

Knots investigation was initially fostered by the discovery of knotted circular single-stranded DNA~\cite{Liu} and has been followed by the study of the underlying enzymatic mechanisms~\cite{Wasserman,Shaw} and more recently by the description of the topological organization and packing dynamics of bacteriophage P4 genome~\cite{Arsuaga,Marenduzzo}.\\
Despite those great advances in knotted DNA studies, we are only beginning to go deeper into protein knots characterization and the understanding of their biological role. After the pioneering work of Mansfield~\cite{Mansfield} and the definition of topological descriptors for the analysis of protein symmetries and proteins classification~\cite{Chen, Emmert, Erdmann},  the detection of knots in proteins was boosted by Taylor's work~\cite{TaylorI}.
 The exponential growth of the total number of structures deposited into the Protein Data Bank (PDB, \href{http://www.pdb.org}{http://www.pdb.org})~\cite{Berman}  requires dedicated computational high-throughput methods able to deal with a large amount of data~\cite{VirnauII}. These methods combine a structure reduction scheme of a protein backbone model with the computation of a knot invariant, the Alexander polynomial~\cite{Chen,Kolesov,Lai,VirnauI}. Hereinafter with the term reduction we refer to a stepwise deletion of a certain number of points from the original structure (endpoints excluded) that preserves its ambient isotopy class.\\
The most affirmed reduction algorithm is the KMT reduction scheme. KMT owes its name to the  different algorithms proposed by Koniaris and Muthukumar~\cite{Koniaris} and Taylor~\cite{TaylorI,TaylorII}. Since the use of this acronym has engendered a little confusion on which algorithm is precisely being used in literature we will explicitly refer to them by  authors' names. Globally, these methods are based on the concept of elementary deformation~\cite{Reidemeister,Alexander}, which consists in the replacement of two sides of a triangle with the third provided that the triangle is empty. In particular while Koniaris and Muthukumar's algorithm essentially reproduces the ideas of Alexander-Briggs and Reidemeister, in the Taylor's algorithm (which Taylor himself considers  a smoothing algorithm) the elementary deformation is done in steps that progressively smooth the chain at the cost of introducing points not belonging to the protein backbone; the edge replacement depends on some selected conditions~\cite{TaylorII} chosen to prevent numerical problems.\\
Once the reduction has been accomplished knot type identification can be performed. This can be done either by visual inspection or by computing a polynomial invariant. Being easy to compute the Alexander polynomial represents the current default choice. This is also supported by the evidence that protein knots detected to date are the simplest ones as illustrated in Figure~\ref{Figure_2}.

Unfortunately, the Alexander polynomial does not distinguish a knot from its mirror image. Thus, for instance left- and right-handed trefoil knots share the same polynomial. Instead, more powerful invariants are able to determine knots chirality.\\
Whereas to define the handedness of the simplest knot types is straightforward, its extension to more complex knots requires carefulness.
However, for the purpose of this article, a knot is chiral if its mirror image and the knot itself belong to two different ambient isotopy classes and it is achiral otherwise. We define the handedness of knots according to~\cite{Liang} adopting the conventional values reported in the Atlas of Oriented Knots and Links (\href{http://at.yorku.ca/t/a/i/c/31.htm}{http://at.yorku.ca/t/a/i/c/31.htm}).\\
As far as proteins are concerned, the handedness of protein knots was only partially addressed so far.\\
Taylor points out the existence of both right- and left-handed trefoil knots, with a neat right-handed preference~\cite{TaylorIII}. This hypothesis was supported by the finding that all trefoil knotted proteins belong to the SCOP~\cite{Murzin} $\beta\alpha$ class, where an intrinsic right-handed preference for $\beta\alpha\beta$ unit connections exists. The only left-handed trefoil knot was detected in the ubiquitin C-terminal hydrolases (1cmx) considered afterwards as an incomplete five crossings knot. However, by considering individual fragments the knot vanishes. A more recent work that removed sequence redundancy, intriguingly highlights a global 5 to 3 balance between right-handed and left-handed knots, not suggesting a bias for one of the two hands~\cite{Micheletti}.\\
In order to compute invariants able to cope with knots chirality, here we propose a novel topological framework to compute arbitrary skein polynomials. A skein polynomial $P$ respects the skein relation:
\begin{equation}
a L_+-b L_-= cL_0
\label{skeinrel}
\end{equation}
which is an algebraic relation connecting the configurations in a Conway skein triple~\cite{Conway} (see Figure~\ref{Figure_1}B), namely it verifies
 \begin{equation*}
 aP(L_+)-bP(L_-)=cP(L_0)
 \end{equation*}
 where the coefficients $a,b,c$ have to satisfy some relations. For instance, the choice $b=a^{-1}$, $c=z$ leads to the HOMFLY polynomial $P(a,z)$~\cite{Freyd}. By further specializing $a=t^{-1}$ and $z=t^{1/2}-t^{-1/2}$ one obtains the Jones polynomial $V(t)$ whereas setting $ a=1$ and  $z=t^{1/2}-t^{-1/2}$ leads to the Alexander polynomial $\Delta(t)$. As far as proteins are concerned, the handedness of protein knots was previously addressed by King et al.~\cite{King} and relies on the computation of the Jones polynomial.\\
Although this appears to be enough to define the chirality of the currently detected knotted proteins, the HOMFLY polynomial is more powerful. For instance, whereas the Jones polynomial is the same for knots 10-022 and 10-035 of the Rolfsen table, the HOMFLY polynomial is able to discriminate them. In the realm of our method, other choices bring to the Vassiliev knots invariants~\cite{Vassiliev,Rogen} considered for instance by~\cite{Lua}.\\
Generally, the skein relation does not preserve the multiplicity of a link. For example if $L_+$ is a knot, $L_0$ will be a two components link. The recursion of the skein relation together with  the values of the given polynomial on the unknot allows to reconstruct the polynomial of any given link. Therefore, the complexity of the polynomial computation grows exponentially with the number of crossings to be processed. Our algorithm relies on the iteration of the skein relation and explicitly constructs the Conway skein triple associated to a given crossing by a stepwise insertion of auxiliary points.\\
In order to deal with multi-component links and speed up computations, the polynomial computation is preceded by the application of a structure reduction scheme, which we call \textsf{MSR} (Minimal Structure Reduction). The \textsf{MSR} algorithm exploits the interplay between the 3D structure and the corresponding 2D planar diagram of a polygonal path and basically relies on a 3D operation, namely the Generalized Reidemeister Move (GRM). While the Alexander-Briggs method intrinsically removes at most one point at each step, a GRM  does not necessarily operate locally, usually leading to a dramatic reduction of the number of points in few steps.\\
The effectiveness and robustness of the proposed framework were initially evaluated on tabulated knots and links, leading to an HOMFLY polynomial repository along with knots orientation details. We then applied our methods to protein structures. By screening the entire PDB (version of November 8, 2010), we obtained an up-to date table of knotted structures that also includes two newly detected right-handed trefoil knots.

\section{Methods}

\subsection{Basic concepts and definitions}
To make this article self-contained, herein we introduce and briefly describe basic concepts and definitions.
\begin{itemize}
\item \textit{Polygonal paths}
A pair $(P,S)$ where $P=\{P_1,\ldots,P_N\}$ is a collection of $N$ points in $\mathbb R^3$  and $S=\{S_0,S_1,\ldots,  S_K\}$ is an ordered subset of $[0\puntini N]$  (the integers  in  $[0,N]$) with $S_0=0, S_K=N$  determines a collection of $K$ polygonal paths in $\mathbb R^3$  as follows:  the $k$-th path (or component) is generated by connecting the points indexed by $(S_{k-1}\puntini S_k]$.\\
The edges of the polygonal paths  are the oriented segments $P_{i}P_{i+1}$ with $i\in E=[1\puntini N-1] \setminus S$.\\
A collection of polygonal paths $(P,S)$ in $\mathbb R^3$  is   simple if each edge of the path intersects precisely the previous and the next edge at the endpoints~\cite{Livingston}.
\item \textit{Polygonal link} A collection $\cL=(P,S)$ of simple polygonal paths is a polygonal link. The $K=K(\cL)$ components of $\cL$ are not necessarily closed. For the sake of convenience, a subpath will be defined by indexing $\cL$ with square brackets.
\item \textit{Regular Projection}
A projection $\pi:\mathbb R^3 \rightarrow \mathbb R^2$  of
a polygonal  link $\cL$ is regular if the following conditions are satisfied:
\begin{enumerate}
	\item The image $\pi(\cL)$ has at most a finite number of double points (crossings).
	\item No vertex is a double point.
\end{enumerate}
A link diagram is a regular projection of the link whose graphical representation adopts solid edges and gaps to indicate overcrossings and undercrossings respectively (see Figure~\ref{Figure_1}A). With a slight abuse of language we will also call under/over crossings the points in $\mathbb R^3$  that project to an over/under crossing in $\mathbb R^2$.
\item \textit{Intersection signs}
Given two sets of edges $A$ and $B$ we can compute the intersection matrix $I=I(A,B)$ by setting 
\begin{equation}
\label{eqintersections}
(I(A,B))_{i,j}=\cases{0 &   if $A_i$ and $B_j$ do not intersect transversally\\
	+1 &  if $A_i$ lays over $B_j$\\ 
	-1 & if $A_i$ lays  under $B_j$\\}
\end{equation}
If $A=B$ we get an antisymmetric square matrix and we can simplify the notation to $I(A)$. Intersection signs definition is detailed in Text~S1.
\item \textit{Minimal structure}
A minimal structure for a polygonal link $\cL$ is a nested sequence of subsets of $\cL$ $$\cL\supset \cL_1\supset\ldots\supset \cL_N$$ that cannot be extended. Each inclusion corresponds to a Generalized Reidemeister Move, described below.
\end{itemize}

\subsection{Structure reduction algorithm}
Our reduction algorithm \textsf{MSR} iteratively exploits the subroutine \textsf{GRM}, which performs a Generalized Reidemeister Move according to the following scheme:
\begin{description}
\item [\rm{Step1:}] Move candidate selection, namely a subpath $\cM$ of $\cL$.
\item [\rm{Step2:}] Move contraction $\cLc$, which is the provisional replacement in $\cL$ of $\cM$ with the segment $\cMc$ connecting the endpoints of $\cM$. 
\item [\rm{Step3:}] Check that $\cL$ and $\cLc$ belong to the same ambient isotopy class. If so, the replacement described in Step2 becomes effective.
\end{description}
While the first two steps are trivial, Step3 requires the study of the intersections of the move candidate $\cM$ with the remainder $C$ of $\cL$. $\cM$ is characterized by its initial and final edge indices, respectively $\bM$ and $\eM$ and belongs to a specified component, say $m$ of $\cL$.\\
The complement $C$ can be splitted in $C_\mathrm{out}$, the link components different from $m$ and $C_\mathrm{in}$, the open link with at most two components given by $\cL[(S_{m-1} \puntini \bM)]$ and $\cL[(\eM+1\puntini S_{m})]$. Let $\mathrm{sign}({\cM})$ be the set of signs of $I(\cM,C)$ and analogously $\mathrm{sign}(\cMc)$ be the set of signs of $I(\cMc,C)$.\\
The topological check in Step3 requires the evaluation of the three following conditions:
\begin{itemize}
\item [] ($T$) $\cM$ is ascending or descending (Triviality of $\cM$).
\item [] ($S$) $\mathrm{sign}({\cM})$ contains at most one element (Separability of $\cM$ from $\cL$).
\item [] ($C$) The set $\mathrm{sign}({\cM})\cup \mathrm{sign}({\cMc})$ contains at most one element (Concordance of $\cM$ and $\cMc$ with respect to $\cL$).
\end{itemize} 
If $TSC$ conditions hold, we call the replacement of $\cM$ with $\cMc$ (and vice versa) a Generalized Reidemeister Move. A GRM is an equivalence relation for polygonal links. An example of an admissible move is illustrated in Text~S2.\\
Given a polygonal link $\cL$, its intersections matrix $I_{\cL}=I(\cL)$ and the move initial index $b$, the \textsf{GRM} algorithm performs the following operations: 
\begin{algorithm}[H]
\algsetup{indent=2em}
\begin{algorithmic} [5]
\STATE Initialize:
\STATE $I_\mathrm{out}=I_\cL$
\STATE $\cL_\mathrm{out}=\cL$
\STATE $e=b+1$
\WHILE{$\left(e \in E \right)$} 
	\STATE $\cM=\cL[[b..e]]$
	\STATE Check Condition $(T)$
	\IF{$(T)$ False}
		\STATE{Go to  \underline{\rm Exit}}
	\ENDIF
	\STATE Check Condition $(S)$ 
	\IF{$(S)$ False}
		\STATE{Goto  \underline{\rm Exit}}
	\ENDIF
	\STATE Compute the vector $r=I(\cMc,\cL)$  
	\STATE Construct $I_{\cLc}$ from $I_\cL$ and $r$
	\STATE Check Condition $(C)$
	\IF{$(C)$ False}
		\STATE{Goto  \underline{\rm Exit}}
	\ENDIF
	\STATE $I_\mathrm{out}=I_{\cLc}$
	\STATE $\cL_\mathrm{out}=\cLc$
	\STATE $e=e+1$
\ENDWHILE
\STATE  \underline{\rm Exit}
\STATE $\cL=\cL_\mathrm{out}$
\STATE  $I_\cL =I_\mathrm{out}$\\
\RETURN{$\cL$ and $I_\cL$}
\end{algorithmic}
\end{algorithm}
The key point of the algorithm is the construction of the intersection matrix $I_\cLc$ from $I_\cL$ (line 16) simply by replacing the rows and columns $[b..e]$ of $I_\cL$ with the vectors $+r$ and $-r$ respectively. Notably, this procedure greatly reduces the computational cost with respect to an explicit matrix computation.\\
We are now ready to introduce \textsf{MSR}. Given a polygonal link $\cL$ and an iteration limit $n$ (suitable to achieve a partial reduction) \textsf{MSR} operates as follows:
\begin{algorithm}[H]
\algsetup{indent=2em}
\begin{algorithmic} [5]
\STATE Compute $I_{\cL}=I(\cL)$
\STATE $l:=\# \cL$ (Dynamic assignment)
\STATE $i=1$
\WHILE{($i\leq n$)} 
	\IF{$l=2\,K$ (where $K$ is the multiplicity of $\cL$)}
		\STATE{Go to  \underline{\rm Exit}}
	\ENDIF
	\STATE $p=\# \cL$
	\STATE $b=1$
	\WHILE{($b<l-1$)} 
		\STATE $({\cL},I_{\cL})\leftarrow\textsf{GRM}({\cL},I_{\cL},b)$
		\STATE $b=b+1$
	\ENDWHILE	
	\IF{$p=l$ (Reached minimal structure) }
		\STATE{Go to \underline{\rm Exit}}.
	\ENDIF
	\STATE $i=i+1$
\ENDWHILE
\STATE  \underline{\rm Exit}
\RETURN{$\cL$}
\end{algorithmic}
\end{algorithm}
%%%

\subsection{Skein polynomials computation}
\begin{figure}
\includegraphics[width=1\textwidth]{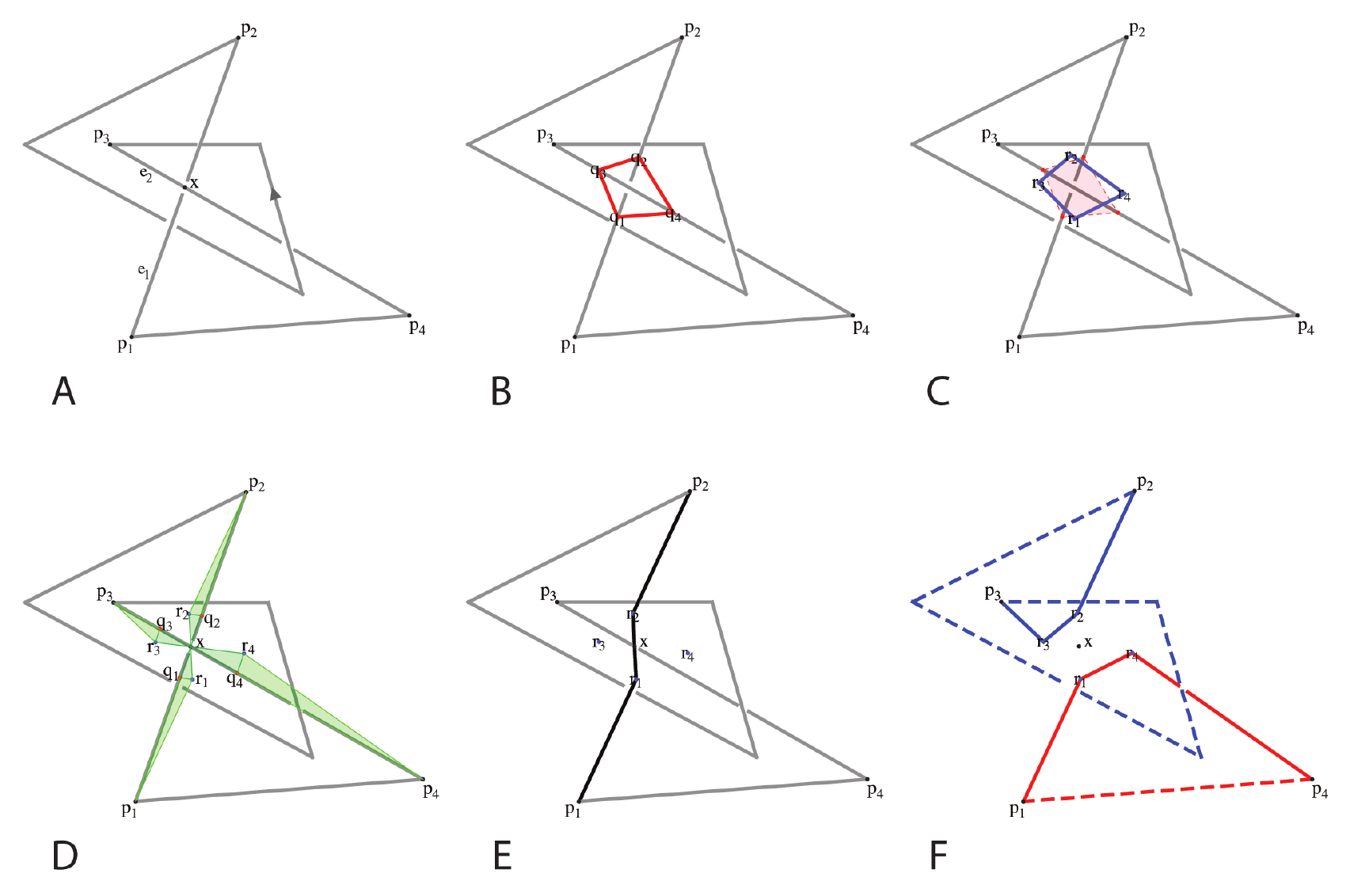}
\caption{
{\bf Example of geometric construction of the skein configurations}  (A) Figure-eight polygonal knot diagram. Knot orientation and the crossing $x$ between the edges $e_1$ and $e_2$ are shown. (B) A clean quadrilateral $q$ around $x$ is shown in red. (C) The rotated quadrilateral $r$ (solid blue lines) is obtained by rotating $q$ (dashed red lines) along the $z$ axis. (D) Triangles to be analyzed in the topological check are shaded in green. The points $q$ and $r$ are reported respectively in red and blue. (E) The $\sw$ configuration, with the path $P_1R_1X_\mathrm{sw}R_2P_2$ highlighted in black (F) The $\cL_0$ configuration. Solid lines highlight new connections $P_1R_1R_4P_4$ (in red) and $P_3R_3R_2P_2$ (in blue).}
\label{Figure_3}
\end{figure}

In the following the interplay between three and two dimensions plays a fundamental role and it is realized through the standard projection $\pi_z$. Since $\pi_z$ restricted to $\cL$ is invertible up to a finite number of double points, we denote with an uppercase letter objects of $\cL$ and with the corresponding lowercase letter their projection. Counter images of double points are distinguished by subscripts. Obviously, any subpath in the projection has a unique lift to $\cL$ and therefore in the following we adopt a two dimensional description.\\
Given a polygonal oriented link, we consider two oriented edges $E_1=P_{1}P_{2}$ and $E_2=P_{3}P_{4}$ such that their projections $e_1=p_1p_2$ and $e_2=p_3p_4$ cross at a point $x$. For the sake of convenience we assume that $E_1$ lays under $E_2$ and we respectively denote by  $X_1$ and $X_2$ their points projecting down to $x$. The edges $e_1$ and $e_2$ give rise to a skein configuration of type $+$~or~$-$.\\ 
We implemented the Skein Relation on the 3D structure of  $\cL$ by construction of the corresponding skein configurations $\sw$ and $\cL_0$. With $\sw$ we refer to the switching of the crossing under consideration. Our algorithm performs the following steps (illustrated in Figure~\ref{Figure_3}):
\begin{description}
\item [\rm{Step1:}] Construct an empty quadrilateral $q$ containing $x$ whose vertices belong to $e_1$ and $e_2$.
\item [\rm{Step2:}] Rotate in 2D $q$ to get $r$ and provisionally change $\cL$ getting $\cL_r$ (by means of the just introduced lift operation).
\item [\rm{Step3:}] Check that $\cL$ and $\cL_r$ are topologically equivalent.
\end{description}

%%Quadrilateral
\subsubsection{Quadrilateral Construction} The edges $e_1$ and $e_2$ are divided in two cut edges by the crossing $x$ (see Figure~\ref{Figure_3}A). We construct a quadrilateral with vertices on the four cut edges such that it contains no other edges of the polygonal link projection (clean quadrilateral, see Figure~\ref{Figure_3}B). We consider the four parametric half lines $r_i$ with  parameters $k_i$, $i\in [1\puntini 4]$ leaving from $x$ along the four cut edges 
\begin{equation*}
 r_i(k_i)= x + k_i(p_{i}-x)
 \end{equation*}
For a given value of the parameter vector $\bf k$ we get vertices of a quadrilateral $q=q(\bf k)$. The vertices follow the order $1,3,2,4$. To construct a clean quadrilateral we proceed as follows:
\begin{enumerate}
\item Initialize ${\bf k}$ by setting each $k_i=0.8$.
\item Construct the quadrilateral $q({\bf k})$ and compute the list of distances $d=\{||q_i-x||\}_{i\in [1\puntini 4]}$.
\item Check the cleanness of $q$ via the \textsf{Xclean} algorithm (described below).
\item If $q$ is not clean, consider $d$ and iteratively reduce by half the parameter associated with the longest cut edge having intersections (which we call $e_\mathrm{max}$). 
\end{enumerate}
%%Xclean function
\textsl{\\ \emph{\textsf{Xclean}} algorithm}
Given an oriented $n$-polygon and a polygonal link we can construct a $n\times 2$ table $S$ of status of the $n$ vertices. Each row of $S$  is a pair summarizing the intersections of the side entering and leaving the vertex as follows: we assign  0 if the relevant side has no intersections with $\cL$ and 1 otherwise.\\
\Xcl \,needs a given quadrilateral $q$, a link projection, a $4\times2$ table $S$ (the putative status list) and a set indexing the vertices whose relevant sides have to be checked. The algorithm simply recomputes the indexed rows of $S$ and updates subsequently the adjacent rows.

%%Rotation
\subsubsection{Quadrilateral Rotation}
As a result of the previous algorithm we end up with a clean quadrilateral $q$, whose vertices lie on $e_1$ and $e_2$. By inserting in $\cL$ the lift of these vertices as auxiliary points we will run into technical problems due to parallel edges. To overcome this problem we generate a new quadrilateral $r$ by rotating $q$ of a suitable angle $\alpha$ around $x$ (Figure~\ref{Figure_3}C) via the the following steps:
\begin{enumerate}
\item Set $\theta=\theta(e_1,e_2)$ equal to  the minimum angle between the vectors $e_1$ and $e_2$.
\item Initialize  
$$\alpha=\mathrm{Min}\left(\frac{\pi}{8},(1-\epsilon)\,\theta\right)$$
where  $\epsilon=0.01$ is chosen such that an edge (e.g. $e_1$) does not bridge the starting position of the other edge (e.g. $e_2$).
\item Construct $r$.
\item Check the cleanness of $r$ through the \Xcl \,algorithm.
\item If not, iteratively reduce by half $\alpha$ until $r$ become clean.
\end{enumerate}
Given $r$ we can construct $\cL_r$ by considering the triangle $p_ir_ix$ (see Figure~\ref{Figure_3}D) and replacing the original cut edges $p_ix$ with the path $p_ir_ix$ (two-side replacement), with \iset. 
%%Topological check
\subsubsection{Topological Check}
The feasibility of the replacement of $\cL$ with $\cL_r$ is not obvious and requires a careful check, which is accomplished analyzing the newly introduced connections.
The triangle $prx$ is subdivided in two triangles by the segment $qr$. The absence of intersections in the segments $qx$ and $rx$ is guaranteed by the cleanness of $q$ and $r$.\\
We approve the two-side replacement if and only if:
\begin{enumerate}
\item The edge $qr$ has no intersections.
\item The segments $pq$ and $pr$ intersect the same edges of $\cL$ preserving intersections order and signs.
\end{enumerate}
Otherwise the rotation angle $\alpha$ is reduced by half and we loop back to Step2. %From now on  $\cL$  will be rigged with the proper quadrilateral $r$ for each crossing.

\subsubsection{Construction of the Skein Configurations}
The construction of the skein configurations requires a distinction between $\sw$ and $\cL_0$.\\
%%Lswitch
To construct $\sw$ we initially take the specular image $X_\mathrm{sw}$ of the undercross $X_1$ with respect to the overcross $X_2$. By replacing the edge $R_1R_2$ with the path $R_1X_\mathrm{sw} R_2$ we obtain a switched crossing but the projection is not regular anymore. Thus, we slightly perturb $X_\mathrm{sw}$ by attracting it toward $R_1$ via the formula
\begin {equation*}
X_\mathrm{sw}\leftarrow R_1+k_\mathrm{sw}(X_\mathrm{sw}-R_1)\hskip50pt  k_\mathrm{sw}<1
\end{equation*}
The constraint on $k_\mathrm{sw}$ guarantees that the projection of $R_1X_\mathrm{sw}$ has no intersections with $\cL$, while the projection $X_\mathrm{sw}R_2$ has one intersection with $e_2$ but it is not always an overpass. If not, we reduce the perturbation via the iterative formula
\begin{equation*}
k_\mathrm{sw}\leftarrow (k_\mathrm{sw}+1)/2
\end{equation*}
whose convergence to 1 guarantees that we will eventually obtain an overpass. We set the initial value $k^0_\mathrm{sw}$ to 0.9.\\
%%Connectionss
Given $X_\mathrm{sw}$, to construct $\sw$ we replace in $\cL$ the edge $P_1P_2$ with the path $P_1R_1X_\mathrm{sw}R_2P_2$ (see Figure~\ref{Figure_3}E). Notice that the edge $P_3P_4$ is not affected by this construction. \\
%%L0
Instead, the construction of $\cL_0$ make a full use of $R$ by substituting in $\cL$ the edges $R_1R_2$ and $R_3R_4$ with the connections $R_1R_4$ and $R_3R_2$ (Figure~\ref{Figure_3}F). Obviously, this determines a shift of the separator indices $S$ and of the numbering of the points following $P_1$. The case where $e_1$ and $e_2$ belong to the same component of $\cL$ is treated differently from the case where they belong to different components. In the former, the number of components of the link increases while in the latter it decreases.
%%%

%%Skein recursion
\subsubsection{Skein recursion}
We will apply recursively the skein relation~(\ref{skeinrel}) to reduce a given polygonal link $\cL$  to a collection of trivial links, systematically switching the undercrossings.\\
We adopt a greedy approach in which at each recursion we switch the undercrossing leading to the $\sw$ structure with the lowest number of points and we accordingly produce the relevant $\cL_0$ configuration.\\
In order to speed up computations, at each step the configurations are reduced with \textsf{MSR}. The resulting structures are stored as nodes in a skein tree, a binary ordered tree rooted at the original link.\\
Our goal is to assign to every node $n$ a pair of weights $(s,P)$ where $s(n)$ is precisely the skein sign of the crossing of $n$ to be switched and $P(n)$ is the link polynomial of $n$. Notice that while $s(n)$ is known, $P(n)$ needs to be computed. We adopt a dynamic bottom-up procedure in which starting from leaves we attach $P(n)$ to inner nodes.\\
Leaves are the simplest nodes since given a leaf $l$, $P(l)$ is known a priori being the polynomial of the $K$-components unlink and there is no undercrossing left ($s(l)=\emptyset$). In the skein tree, every inner node $\cL$ has two children, say $\sw$ and $\cL_0$, and $P(\cL)$ can be computed via the recursion formula
\begin{equation*}
 P(\cL) =\cases{a^{-1}b \cdot P(\sw)+ a^{-1}c \cdot P(\cL_0) &if \; $s(\cL)=+1$\\
b^{-1}a \cdot P(\sw)- b^{-1}c \cdot P(\cL_0)&  {if}\; $s(\cL)=-1$}
\end{equation*}
In this way, the polynomial is simply the weight $P$ of the root.

% Results and Discussion can be combined.
\section{Results and Discussion}
\subsection{Validation on tabulated knots and links}
\begin{figure}
\includegraphics[width=0.8\textwidth]{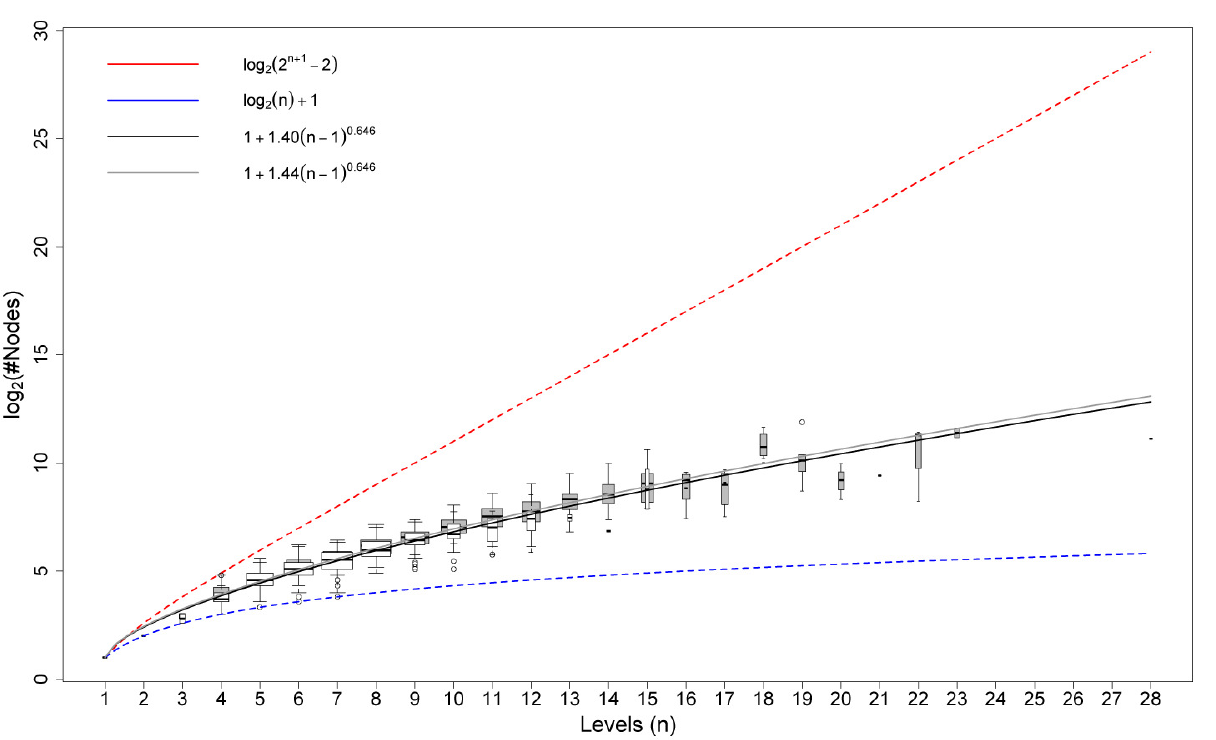}
\caption{
{\bf The Increase of the number of tree nodes as a function of tree levels.}  Trees of both greedy (white/black) and fixed choice (gray) algorithms have been clustered according to the number of levels ($n$). For each cluster a box plot of the nodes number has been drawn with a width proportional to the cluster size.  Solid power curves fit the reported data. Dashed red and blue curves represent respectively lower and upper estimates of node numbers. Curve expressions are shown in the legend.}
\label{Figure_4}
\end{figure}

Initially, we validated our methods by computing the HOMFLY polynomial of both full structures and minimal stickies representations of tabulated polygonal knots and links. We compared our results with a polynomial repository constructed as described in Text~S1. Since standard repositories do not address orientation and chirality, a single polynomial is associated to a given structure and a computed polynomial could not directly match repository entries. Thus, for each tabulated structure we considered mirror images along with all possible orientations (together referred to as flips) and computed the corresponding polynomials. At least one of them matched the one reported in the polynomial repository.
Our complete repository of knots up to 10 crossings and oriented links up to 4 components could be browsed at \href{http://www.pharm.unipmn.it/rinaldi/knots/index.php}{http://www.pharm.unipmn.it/rinaldi/knots/index.php}.\\
As described above, our HOMFLY polynomial computation associates a skein tree to every knot or link, by means of a greedy selection of the crossing to be switched. To verify the goodness of this choice we compared it with a fixed choice variant, which systematically switches the first -1 crossing encountered. We applied both algorithms to every knotted structure in the repository (including flips), characterizing each tree with two complexity indices, namely the level (corresponding to the number of generations, $n$) and the number of tree nodes $k$. Figure~\ref{Figure_4} shows the behavior of $k$ as a function of $n$, with dashed curves representing theoretical constraints. The growth curves of the two algorithms obtained via ANCOVA after linearization are significantly different, showing that the greedy algorithm performs generally better than the fixed choice one. This result is also supported by the evidence that the number of levels and configurations required for polynomial computation is significantly lower for the greedy choice (Wilcoxon test on the pairwise differences, $p<10^{-15}$). Notably, the shrinking of the tree well compensates the extra computational time required by the greedy choice and this particularly suggest the usage of this algorithm as structure complexity increases. In general, it is possible to find a time threshold such that by filtering computational times accordingly, a significant difference emerges supporting greedy choice. This suggested the adoption of the greedy algorithm for the reduction of protein structures.

\subsection{Application to protein structures}
\begin{figure}
\includegraphics[width=0.8\textwidth]{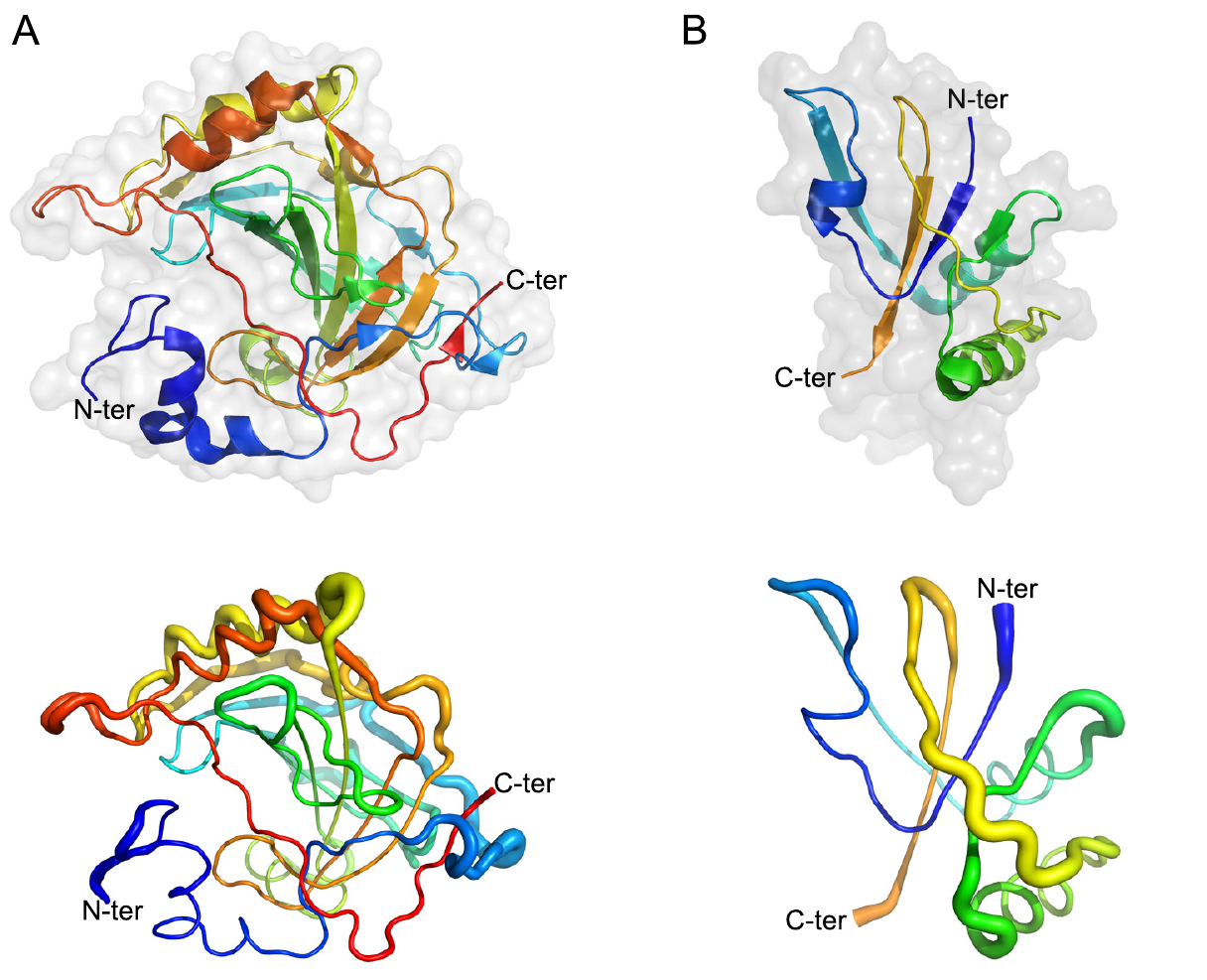}
\caption{
{\bf The two newly identified right-handed trefoil knots in recently deposited protein structures} (A) On the top, the secondary structure and the accessible surface area (in transparency) of the human Carbonic Anhydrase VII, isoform 1 (3mdz) is shown. On the bottom, a sausage view cartoon of the same enzyme is shown. In this representation, the diameter of the sausage is proportional to the B-factor. The thicker the backbone is, the more flexible it is. (B) The same representations as in (A) are shown for the knotted core of the uncharacterized ORF from \textit{Sulfolobus Islandicus rudivirus 1} (2x4i), chain A.
Colors change continuously from blue (first residue) to red (last residue). The last residue of the 2x4i protein is colored in orange, since the structure presents a gap toward its true C-terminus end and results a slip-knot when the whole structure is considered, as detailed in the text.
}
\label{Figure_5}
\end{figure}

We applied our algorithms to all the protein structures deposited in the PDB. Each entry was preprocessed as described in Methods and the HOMFLY Polynomial was computed on the \textsf{MSR} reduced structures.\\ 
Globally, we found 119 knotted proteins (226 parts) of the five knot types shown in Figure~\ref{Figure_2}, belonging to the ten previously well defined classes of knotted foldings~\cite{VirnauII,Micheletti}. A summary table of knots for each knot type along with the relevant HOMFLY polynomial is reported in Table~1. For a complete list of knotted proteins ID and part details see Table~S1.\\
\begin{table}[!h]
\small
\caption{
\bf{Total knotted entries detected for each knot type.}}
\begin{tabular}{|c|c|c|c|c|}
\hline
knot type & handedness  & \#structures & \#parts & HOMFLY polynomial\\
\hline
        $3_1$ & R &103 & 184 & $-l^{-4}+2l^{-2}+l^{-2}m^2$\\
        $3_1$ & L &3 & 3 & $-l^4+2l^2+l^2m^2$ \\ 
        $4_1$ & - &10 & 31 & $-1+l^{-2}+l^2-m^2$\\ 
        $5_2$ & L &2 & 4 & $l^2+l^4-l^6+(l^2+l^4)m^2$\\
        $6_1$ & R & 1 & 4 & $l^{-4}-l^{-2}+l^2-(1+l^{-2})m^2$\\ 
\hline
\end{tabular}
\begin{flushleft} Entries show the number of knotted structures and relevant parts for each knot type.
\end{flushleft}
\end{table}
Although redundancies with previous studies~\cite{VirnauII,Lai,Micheletti} are largely present, the number of knotted proteins is lower than what previously reported. This is mainly due to topological checks and distance controls (see also Text~S1) that allowed to discard nonstandard PDB formats and entries having large structural gaps due to missing residues. These proteins are often detected as knotted when gaps are connected by straight lines, inducing artificial entanglement.\\
Among newly detected knotted proteins, two right-handed trefoil knots were identified in two recently deposited structures. The first one has been found in the human Carbonic Anhydrase VII (CA7), isoform 1 (3mdz) (see Figure~\ref{Figure_5}A), whereas second one has been detected in the uncharacterized ORF from \textit{Sulfolobus Islandicus rudivirus 1} (2x4i) (Figure~\ref{Figure_5}B), a virus of the extremely thermophilic archaeon \textit{Sulfolobus}. Notably, although the latter protein still needs to be fully characterized to define its relevance, it shares more than 50\% of its primary sequence with protein B116 (2j85)  of \textit{Sulfolobus turreted icosahedral virus}, which King et al~\cite{King} previously reported to contain a slip-knot. Thus, it is not surprising that the structure of 2x4i also contains a slip-knot, as we confirmed by visual inspection. Moreover, this protein presents a gap toward its C-terminus. Since we treat gaps as chain terminators (see Text~S1) what we have detected is the knotted core of the slip-knot, illustrated in Figure~\ref{Figure_5}B. The trefoil knot in the CA7 belongs instead to the well known right-handed trefoil knotted Carbonic Anhydrase superfamily. Knotted core analysis, performed as reported in~\cite{TaylorI,VirnauI}, reveals that both knots have a quite shallow nature. While a trimming of 28 and 5 residues from the N-terminus and C-terminus respectively is sufficient to unknot the Carbonic Anhydrase VII, the uncharacterized ORF becomes unknotted after an even deletion of 5 residues. However, this is sufficient to exclude an artifactual nature of these knots.\\
For what concerns recently reported trefoil knots, our results confirm the presence of a right-handed trefoil knot in the alpha subunit of human S-adenosylmethionine synthetase 2 (2p02) and the artifactual origin of the one detected in the ribosomal 80S-eEF2-sordarin complex of \textit{Saccharomyces cerevisiae} (1s1h) first reported in~\cite{Micheletti}.\\
Interestingly, we detected three left-handed trefoil knots respectively in the U2 snRNP Rds3p protein of \textit{S. Cerevisiae} (2k0a), VirC2 protein of \textit{Agrobacterium tumefaciens} (2rh3) and in the uncharacterized protein MJ0366 from \textit{Methanocaldococcus jannaschii} (2efv). A fourth knot detected in the human prothrombin complexed with a peptidomimetic inhibitor (1jwt) was discarded due to a long structural gap. The left-handed trefoil knot in the Rds3p protein, which highlight a knotted zinc-finger motif, is the deepest knot of this kind reported to date~\cite{vanRoon}. Indeed, its knotted core is preserved after trimming of 19 and 18 residues from the C-terminus and the N-terminus respectively. Since this protein does not resemble protein belonging to the $\beta\alpha$ class, it shifts the left-handed to right-handed balance to 4 to 5, thus enforcing the non preferential handedness hypothesis.

\subsection{Analysis of the \textsf{MSR} algorithm}
\begin{figure}
\includegraphics[width=0.8\textwidth]{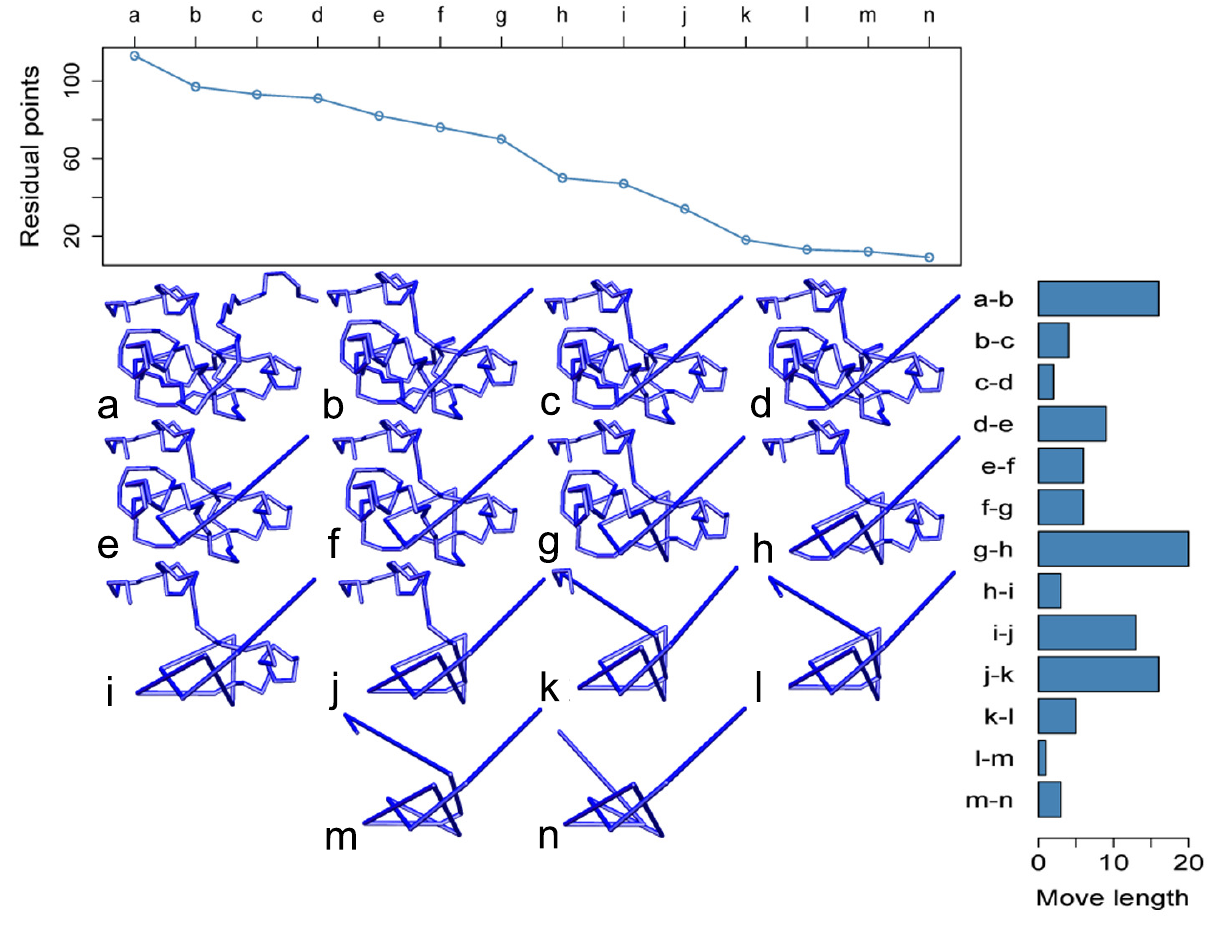}
\caption{
{\bf \textsf{MSR} reduction curve of the U2 snRNP protein Rds3p}  On the middle are illustrated the 13 reduction steps (b-n) for the Rds3p protein (2k0a) (a). The last frame (n) represents the minimal structure of the protein, a left-handed trefoil knot. On the top, the residual points are plotted for each frame a-n. The corresponding move lengths are shown on the right.}
\label{Figure_6}
\end{figure}

\begin{figure}
\includegraphics[width=1\textwidth]{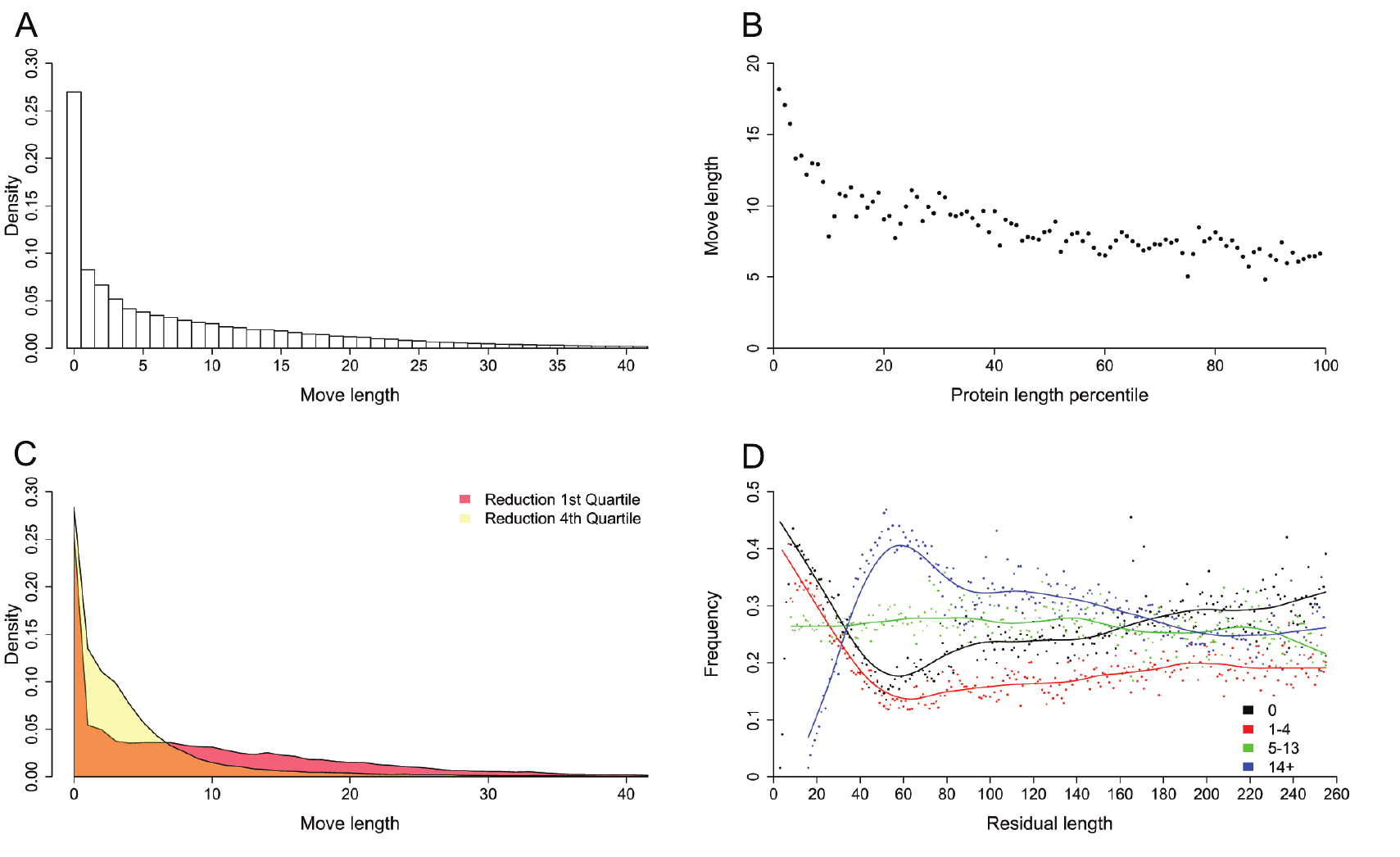}
\caption{
{\bf \textsf{MSR} algorithm analysis}  (A) The observed distribution of move lengths, considering only density values greater than 0.2\%. (B) The mean move length significantly decreases as a function of the protein length. Values for each length percentile are reported. (C) Move length distributions are shown relatively to the first and fourth quartile of the reduction process, considering only density values greater than 0.2\%. (D) Frequencies of classes of move lengths as a function of protein residual length at which they occur are dotted. LOESS curves are reported. Classes cutoffs were chosen according to move length quartiles (0,4,13) and the last 5\% of residual lengths were discarded to remove frequency fluctuations.}
\label{Figure_7}
\end{figure}

As a secondary goal, we were interested in the characterization of an intrinsic feature of the \textsf{MSR} algorithm, the move lengths. Remarkably, differently from other proposed reduction schemes, here the move length is not constrained a priori to one (this can be easily seen in the animated reduction provided as Video~S1). This characteristic leads to a particularly interesting class of curves which we call reduction curves, representing the time series of residual points during the reduction process. For example, Figure~\ref{Figure_6} illustrates the reduction of the above mentioned U2 snRNP Rds3p, the relevant reduction curve and move lengths.\\
To analyze these two features, 19316 protein structures were randomly extracted from the PDB, further selecting only those proteins of length comprised between the first (37 points) and the ninth deciles (357) of protein lengths (15529 structures). Proteins were processed with \textsf{MSR} and the number of residual points was associated to the corresponding move length at each reduction step.\\
We first analyzed moves distribution. The observed distribution of move lengths is shown in Figure~\ref{Figure_7}A, showing that quite long moves are rather frequent. In particular, move lengths quartiles are 0,4,13, the mean is 8.61 and 27\% of the moves have length 0. \\
We then tested if move length depends on protein length. Proteins were sorted by length and the relevant move lengths were grouped in 100 equal sized bins, so that for instance the first bin contains moves corresponding to shortest proteins. As shown in Figure~\ref{Figure_7}B, the mean of each bin significantly decreases (Mann-Kendall trend test, $p<10^{-15}$) as a function of the protein length. An effect of final moves has been excluded by considering only the first 90\% of the reduction process.\\
To assess if move length distribution changes during structure reduction, we compared the move distributions of the first and fourth quartile of the reduction process. To avoid overlaps, we considered reduction sequences of length at least 4 (14346 sequences). A significant difference between the two quartiles emerged (Wilcoxon test, $p<10^{-15}$), as highlighted in Figure~\ref{Figure_7}C. Moves with length up to 6 (short moves) are more frequent toward the end of the reduction process, while long moves occur preferentially in the first reduction quartile. This behavior is also confirmed by comparing the first and second half of the reduction process. However, shorter final moves are in principle explained by an increase of the edges mean length, as can be seen in Figure~\ref{Figure_6}.\\
Finally, an interesting effect emerges when the frequencies of move lengths were analyzed as a function of the residual protein lengths at which they occur. By grouping move lengths in quartiles, while moves below the median reach the minimum frequency for a residual length around 60, the opposite behavior is attained by moves above the median (Figure~\ref{Figure_7}D). Interestingly, a residual length around 60 is the optimum of the reduction process, where the frequency of 0 moves reaches its minimum and contextually the frequency of long moves is maximum.

\subsection{Running time and complexity}
The computation of the HOMFLY polynomial is known to be NP-hard~\cite{Jaeger,Chen} and its running time exponentially increases with the number of crossings in the projection.
However, the application of the \textsf{MSR} algorithm before the polynomial computation dramatically reduces the number of crossings, leading to a feasible computation of the HOMFLY polynomial for any structure analyzed in the present work. Indeed, the \textsf{MSR} algorithm has complexity $O(N^2)$ in the number of points (i.e. the number of residues for a protein) and represents the dominant term in the total computational time for the vast majority of the analyzed structures, often independently from their knotted nature.\\
In practice, running times are reasonable for any analyzed PDB entry on a 2.4 GHz Intel Core 2 Duo processor with 2 Gb of RAM. On average, proteins of length 100, 200 and 300 take respectively 2, 10 and 20 seconds to be processed. The identification of the left-handed trefoil knot in the Rds3p (2k0a) requires 2.8 seconds (2.5 seconds for the \textsf{MSR} algorithm + 0.3 seconds for the polynomial computation), whereas the processing of the Stevedore's knotted protein (3bjx) takes 23.5 seconds (20 seconds + 3.5 seconds).

\subsection{Implementation}
All code for this work was written in Wolfram Mathematica 7 and executed on a Mac OSX platform. We developed the Mathematica package \texttt{HPKnots.m} based on the code provided as Text~S3. \texttt{HPKnots.m} can be obtained upon request. The validation code also required \texttt{KnotTheory.m}, a third-party Mathematica package~(\href{http://katlas.org}{http://katlas.org}).

\subsection{Conclusions}
We have presented a novel topological framework for the HOMFLY polynomial computation of polygonal paths based on the geometric construction of Conway skein triples. Validation on tabulated knots and links demonstrates the global method robustness and the effectiveness of the greedy selection of the crossing to be switched. These evidences have been further confirmed by the polynomial computation of protein structures, also leading to an up-to date table of knotted structures. Whereas the performed topological checks allowed to discard artificially entangled proteins, two new right-handed trefoil knots have been detected.\\
Remarkably, the application range of the presented framework is not limited to proteins and it can be extended to the topological analysis of biological and synthetic polymers. Particularly, the study of knotted synthetic polymers like polyethylene has led to insights into the mechanical properties of such structures. The presence of a knot strongly weakens the polymer that potentially breaks at the entrance to the knot. Furthermore, knots frequency depends on the solvent and is higher in the coil phase than the globular phase with the knotted core size that increases as a function of the number of monomers. These aspects have been previously addressed with the computation of the Alexander polynomial in numerical simulations based on a simplified model of polyethylene~\cite{VirnauI}. Our  framework can be successfully applied to this model and possible refinements, contributing to extend the knots spectrum so far considered and providing information about the knots chirality. Another suitable field of application of our method, in which generally more complex knots are investigated, is the topological study of cyclized DNA~\cite{Shaw,Arsuaga,Marenduzzo}. \\
Finally, the applicability of the presented method is not confined to single component structures and can be applied to the topological study of multicomponent polygonal paths, providing a robust identification of knots or links when the frequency of entangled structures has to be addressed.

\section*{Acknowledgments}
We are grateful to Prof. Giovanni Gaudino for his support. We also thank Prof. Giovanni Battista Giovenzana and Dr. Franca Rossi for a lot of fruitful discussions and insights. Finally, we thank the reviewers for their helpful suggestions on the manuscript.

\newpage
\section*{References}
% The bibtex filename
\bibliographystyle{iopart-num}
\bibliography{comoglio_rinaldi_PLOS_ArXiv}

\end{document}